\documentclass[journal]{IEEEtran}

\ifCLASSINFOpdf
\else
\fi
\usepackage{amsmath}
\usepackage{amssymb}
\usepackage{graphicx}
\DeclareGraphicsExtensions{.pdf,.png,.jpg,.tif,.bmp}
\usepackage{multirow}
\usepackage{multicol}
\usepackage{array} 
\usepackage{threeparttable}
\usepackage{booktabs}
\usepackage{url}
\usepackage{hyphenat}
\usepackage{footnote}
\usepackage{subfigure}
\usepackage{epstopdf}
\usepackage{epsfig}
\usepackage{soul}
\usepackage{cite}

\usepackage[dvipsnames]{xcolor} 
\usepackage{hyperref}

\begin{document}

\title{Mitigatiton of H.264 and H.265 Video Compression for Reliable PRNU Estimation}

\author{Enes~Altinisik, Kasim Tasdemir, Husrev~T.~Sencar
\thanks{Enes~Altinisik is with TOBB University, Ankara, Turkey.
Kasim Tasdemir is with Abdullah Gul University, Kayseri, Turkey.
H. T. Sencar is with Qatar Computing Research Institute, HBKU, Qatar and TOBB University, Ankara, Turkey. 
}
}
\maketitle

\begin{abstract}

The photo-response non-uniformity (PRNU) is a distinctive image sensor characteristic, and
an imaging device inadvertently introduces its sensor's PRNU into all media it captures.
Therefore, the PRNU can be regarded as a camera fingerprint and used for source attribution.
The imaging pipeline in a camera, however, involves various processing steps that are detrimental to PRNU estimation.
In the context of photographic images, these challenges are successfully addressed and the method for estimating a sensor's PRNU pattern is well established.
However, various additional challenges related to generation of videos remain largely untackled.
With this perspective, this work introduces methods to mitigate disruptive effects of widely deployed H.264 and H.265 video compression standards on PRNU estimation. 
Our approach involves an intervention in the decoding process to eliminate a filtering procedure applied at the decoder to reduce blockiness. 
It also utilizes decoding parameters to develop a weighting scheme and adjust the contribution of video frames at the macroblock level to PRNU estimation process. 
Results obtained on videos captured by 28 cameras show that our approach increases the PRNU matching metric 
up to more than five times over the conventional estimation method tailored for photos. 



\end{abstract}

\section{Introduction}

Source attribution is a crucial task in digital forensics that deals with the problem of establishing reasonable certainty about the source of an evidence.
This is a very challenging task as it rests on uniqueness of source objects and requires methods that can distinguish traces made by a particular source object from traces made by every other object.
Among various kinds and sources of digital evidence, such unique characteristics are hard to find.
Fortunately in digital imaging domain, this challenge has been successfully met by demonstrating that photo-response non-uniformity (PRNU) of an imaging sensor can be used for attribution purposes.

Today, it is well established that the this unique and stable noise-like pattern arising due to manufacturing variations in imaging sensors can be used to reliably identify an imaging sensor. 
In practice, this means multimedia content captured by a digital camera, such as photos and videos, can be attributed to their source cameras.  
The development of this capability has so far been focused on photographic images with very limited interest in videos. 
Given videos are becoming increasingly prevalent on the Internet and digital forensic practitioners are just as likely to encounter videos as much as photos in user devices, it is very important that the same attribution capabilities are also available for videos. 

In effect, the PRNU of a sensor superimposes a distinctive noise-like pattern that is modulated with the light intensity on to the sensor output image.
Hence, the ability to attribute a media to its source essentially relies on correct estimation of this pattern.
In any camera, however, raw image data captured by the sensor must be processed in steps through an imaging pipeline before a photo or video is created. 
Therefore, PRNU noise pattern of a sensor needs to be extracted from the camera output media which is crucially a processed and encoded version of the raw sensor output. 
These processing steps in a digital camera pipeline have important implications on how extraction should be performed and on the reliability of PRNU pattern matching. 
Not only they may cause significant estimation errors in the PRNU noise pattern, but they may also introduce artefacts that yield spurious similarities between distinct PRNU patterns.
Although the steps involved in generation of a photo greatly overlaps with those of a video, video generation steps are much more involved and disruptive to PRNU noise pattern estimation.

When recording a video, a camera measures the visible light for a period of time and transform it to a digital data stream.
During this transformation, there are several stages that could suppress the PRNU pattern overlaid in each sensor image. 
At the image acquisition stage, an indispensable processing step is the downsizing of the full-frame sensor output. 
To reduce the amount of data that needs processing, cameras deploy downscaling, cropping, or pixel binning as mechanisms for resolution reduction.
Another key processing step employed by all modern day cameras is the image stabilization which compensates for the effects of camera shake. 
This typically involves application of global and local geometric transformations to align consecutive images. 
These are followed by processing steps, such as white-balancing, demosaicing, noise-reduction, and color transformation which collectively make up the imaging pipeline and are utilized during acquisition of both photos and videos.
Involved processing steps at this stage are expanded with more sophisticated capabilities as cameras' computational power increases.


The second stage is the video compression. 
A raw video is a sequence of images; therefore, its size is impractically large to store or transfer. 
Video compression exploits the fact that frames in a video sequence are highly correlated in time and aims at reducing the spatial and temporal redundancy 
so that as few bits as possible are used to represent the video sequence. 
Consequently, compression related information loss causes much more severe artefacts in videos as compared to those in photos. 
Currently, the most prevalent and popular video compression standards are H.264/AVC and its recent descendent H.265/HEVC.


In essence, successful estimation of the PRNU of an imaging sensor from a given video requires taking these additional processing steps into account. 
With this motivation, in this work, we focus on the impact of video compression on source camera identification and introduce how to deal with adverse effects of H.264 and H.265 compression standards.
We note here that among all the processing steps in the video pipeline, electronic stabilization is the most detrimental to PRNU estimation.
Nevertheless, there are still many cases where stabilization is not deployed when capturing a video.
For example, smartphones running Android and iOS do not support stabilization in their front cameras. 
In addition, Android operating system supports video stabilization when the video resolution is under $1920 \times 1080$ pixels 
and does not guarantee its deployment if the frame rate is above 30 frames/second \cite{androidDev}.
Similarly, iOS allows stabilization only at certain frame resolutions and frame rates depending on the device \cite{apple}.
Regardless of whether stabilization is performed or not, the ability to effectively estimate PRNU from any video requires dealing with compression related artefacts as it is always performed.
The contributions of this work are twofold. 
First, we address disruptive effects of a filtering function deployed by video codecs to suppress coding related visual artefacts.
We describe how one should intervene in the decoding process to minimize these effects rather than seeking to compensate them in post-processing.
Second, we introduce a method to cope with compression related information loss. 
Our approach exploits the abundance of block-level data by either masking out heavily compressed blocks or weighting PRNU contribution of each block in accordance with the amount of compression applied to each block. 
The effectiveness of the proposed methods in PRNU estimation are validated on a custom video dataset obtained from 28 smartphone devices using a camera app we built for Android smartphones to capture videos in a controlled manner.  

The remainder of this paper is organized as follows. 
In the next section, we describe the PRNU estimation process tailored mainly for photographic images and review the current approaches towards extending this capability to videos.
Section III discusses critical steps of H.264 and H.265 video coding from a perspective of how they interfere with PRNU estimation. 
Section IV provides details of our approach in dealing with video coding. 
Experimental results and our conclusions are given in Sections V and VI, respectively.

\section{PRNU Estimation from Still Images and Extension to Videos}

The PRNU is essentially a systematic noise caused by the variation among pixels of an imaging sensor in their sensitivity to the light. 
In \cite{chen},  Chen {\em et al.} provided a mathematical model to characterize this intrinsic variability.
According to this model, the raw sensor output can be expressed in matrix notation as 
\begin{equation}
I=I^{(0)} + I^{(0)} \times K + \theta
\label{eq1}
\end{equation}
where $I^{(0)}$ is the intensity of incident light on the sensor, $K$ is a constant matrix encompassing pixel sensitivities with values distributed around 1, $I^{(0)} \times K$ is the componentwise multiplication corresponding to the PRNU, and $\theta$ is a combination of all other random noise components.
Since $K$ is the multiplicative factor that gives rise to PRNU overlaid to all sensor outputs, it also serves as a unique identifier for the sensor.  
The factor $K$ is estimated through a maximum likelihood procedure using a set of images, $I_{1},\ldots,I_{N}$, as 
\begin{equation}
K=\dfrac{\sum_{i=1}^{N} I_{i} \times W_{i}}{\sum_{i=1}^{N} (I_{i})^2}
\label{eq2}
\end{equation}
where $W_{i}$ represents the noise residue obtained after denoising image $I_{i}$  to suppress image content's interference on the estimation. 

To determine whether a given image $I_{q}$ is captured by a sensor with an estimated PRNU factor $K$, a detection statistic based on normalized correlation of the noise residue $W_q$ and the estimated PRNU $I_q \times K$ is used. 
It has been, however, empirically observed that sensors yield PRNU noise patterns of varying strength which in turn requires adjusting a detection statistic for each sensor. 
To address this problem, Goljan {\em et al.}  \cite{goljan2008digital} introduced the use of peak-to-correlation energy (PCE) as a measure.
The PCE essentially measures the percentage of the total power in the correlation plane that is concentrated in the correlation peak, and due to this normalization it yields a largely sensor-independent detection statistic. 

This estimation and detection process obviously leaves out many details related to subsequent processing in the camera pipeline. 
In reality, the sensor output $I$ undergoes several processing steps, such as color interpolation, filtering, color correction, and lossy compression, before it can be analyzed.
Further the design and operation of the sensor at the hardware level may introduce additional complexities.  
The most critical aspect of these out-of-model factors relates to whether they introduce any systematic artefacts as they may lead to false correlations during detection.
Therefore, in the presence of such artefacts, the estimated PRNU has to be further processed to eliminate them.

In the case of photos, the estimation process has been adapted to combat with processing related artefacts in two main ways \cite{chen}.
The first one concerns with removal of biases introduced to PRNU estimate mainly by the demosaicing operation. 
The differences in offset gains applied to each color component at the sensor output and the periodic nature of color filter arrays together superimpose a periodic pattern onto interpolated color values. 
To eliminate this pattern, row and column averages of the estimated PRNU factor $K$ are successively removed from each element of $K$. 
The second measure involves removal of less deterministic artefacts such as the blockiness artefact caused by the JPEG compression
To suppress this, Fourier domain representation of $K$ is Wiener filtered, and the noise component that remains after the removal of all structural noise patterns in $K$ is used as the sensor identifier. 

As the sensor operation during acquisition of still photos and videos remains unchanged, the basic model in Eq. \ref{eq1} is valid for both types of media.
However the processing steps involved in a video pipeline are more complex and add further complications to reliable attribution of a video to its source.  
Therefore, the process for estimating the PRNU factor $K$ has to be adapted to tackle additional artefacts related to video acquisition.
Several approaches have been proposed to extend this capability to videos. However, reliable estimation of PRNU factor $K$ from videos remained largely an unmet challenge.

The majority of the work in this domain has been concerned with video compression as it is typically more lossy than image compression, and hence more detrimental to PRNU estimation. 
The earliest work, \cite{PRNUvideoJessica}, exploring this problem targeted artefacts, such as blocking and ringing noises, that are more apparent in video frames due to macroblock level operation of video coding.
Assuming most codecs use $16 \times 16$ sized macroblocks, a frequency domain method is proposed for removing signal components that leak into the estimated $K$ exhibiting this periodicity pattern.
A significant limitation of this approach is that, when coding a video frame, codecs use adaptive makroblock sizes that start from $4 \times 4$ and go up to $64\times 64$ while also supporting rectangular sized blocks.
This versatility not only makes identifying a periodicity pattern difficult but the subsequent filtering step results with further weakening of the PRNU pattern.
To better suppress compression artefacts, \cite{MACEfilter} proposed an alternative approach utilizing a minimum average correlation energy (MACE) filter which essentially minimizes the average correlation plane energy, thereby yielding higher PCE values.
By applying MACE filter to the estimated PRNU factor $K$, 10\% improvement in identification accuracy is obtained over the approach of \cite{PRNUvideoJessica}.



In \cite{analiz}, the authors studied the effectiveness of the PRNU estimation and detection process on singly and multiply compressed videos.
For this purpose, videos captured by webcams are compressed using a variety of codecs at varying resolutions and matched with the reference PRNU pattern estimated from raw videos.
Their results highlight the dependency on the knowledge of encoding settings for successful attribution and the non-linear relation between quality and the detection statistic.
Later, \cite{IPBdiff} explored the reliability of the use of different types of frames in PRNU estimation. 
They found that intracoded frames ({\em i.e.}, I frames) yield better results as compared to predicted frames ({\em i.e.}, P and B frames) and suggested giving a higher weight to contribution of intracoded frames during estimation.  
This finding was essentially a revelation of the fact that I frames typically undergo a lighter compression than other types of frames.

To better deal with compression artefacts, \cite{eusipco18} proposed the idea of modifying the operation of the video decoder and utilizing the macroblock level information.
This represents a more active approach to suppressing artefacts than solely relying on post-estimation processing. 
Results of this work showed that turning-off the deblocking filter and eliminating the heavily compressed blocks
promises significant improvement in the accuracy of source attribution. 

Unlike the video compression aspect, only few works addressed the problem of source camera attribution for stabilized video \cite{taspinar2016, luisapaper}. 
Assuming video frames have undergone some global affine transformation during electronic stabilization, these work in common perform a brute force search for the unknown parameters ({\em i.e.}, for scaling, rotation, and shift) that enable proper alignment of PRNU estimates of individual frames. 
In \cite{taspinar2016}, authors proposed aligning all PRNU patterns estimated from I frames with respect to a reference frame prior to estimating the PRNU factor $K$ of the camera sensor. 
In contrast, \cite{luisapaper} proposed first evaluating all pair-wise matches and identifying group of frames that are transformed similarly and then re-aligning the PRNU estimates from each group with respect to each other to obtain the factor $K$.  
These works and others pointed out that downsizing operation in camera pipeline needs to be taken into account when verifying the source of a video 
and considered coping with it through search and re-scaling. 
To address this gap, in \cite{ei2019}, the downsizing behavior of more than 20 cameras is examined, and the effects of in-camera downsizing on accuracy of source attribution is quantified.
This work also introduced a systematic approach for matching PRNU estimates obtained from media acquired at different resolutions.

In this work, we essentially expand on our approach initially introduced in \cite{eusipco18} to cope with video compression with further improved methods while considering both H.264 \cite{marpe2006h} and H.265 \cite{sullivan2012overview,sze2014high} coding standards.
Since video compression is the last processing step in the camera pipeline, deployment of this approach is necessary to tackle stabilization in the case of moderate to heavily compressed videos. 
In the next section, we provide an overview of video compression steps with an emphasis on steps that are most disruptive to PRNU estimation before presenting our solution.

\section{H.264 and H.265 Video Coding Standards}

Video and image compressions are fundamentally different.
A video is a sequence of highly correlated pictures.
This redundancy is reduced by transferring an image region once, then letting the receiver construct other similar image regions using the received reference image and the prediction parameters.

\begin{figure}[ht!]
\centering
\includegraphics[width=0.7\columnwidth]{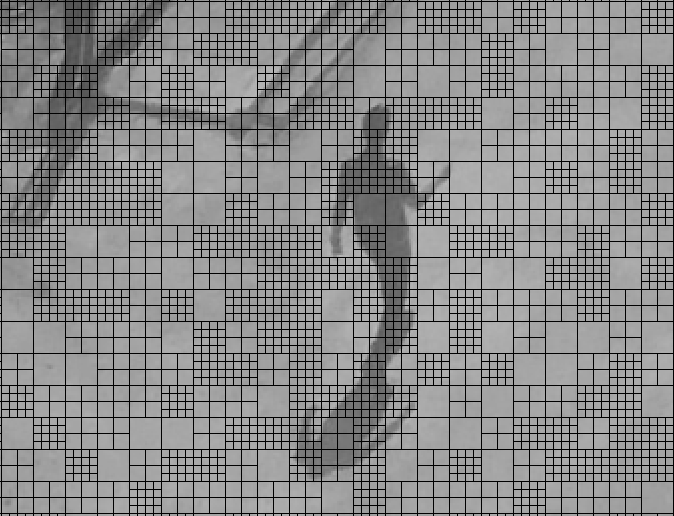}
\caption{A typical Partitioning of an H.264 video frame obtained using Elecard software tool.}
\label{fig:Fig1}
\end{figure}

Similar to JPEG image compression standard, H.264 and H.265 video compression standards divides an image or frame into smaller blocks.
Each coding standard has different naming and size limits for those blocks. In H.264, the largest blocks are called macroblocks and they are of size $ 16 \times 16 $ pixel. A macroblock can be further divided into smaller sub-macroblock regions as small as $4\times4$ pixels. (Figure \ref{fig:Fig1} shows partitioning of a sample video frame in H.264 format.)
H.265 format is an improved version of H.264 and it is designed with parallel processing in mind. The sizes of container blocks in H.265 can vary from $ 64 \times 64 $ to $ 16 \times 16 $.
They might contain smaller sub-blocks as small as $ 4 \times 4 $ pixels.
Unlike the previous format, H.265 allows sub-blocks to have their own sub-blocks which results in a block tree.
Moreover, there are several block types such as Coded Block (CB), Transfer Block (TB) and Prediction Block (PB).
Block definitions and limitations in H.265 is more intricate than the previous version.
However, those details are unrelated to PRNU and skipped for the sake of brevity.
Throughout this paper, related block type of H.265 is referred as macroblock even though the term macroblock does no longer exist in the H.265 standard.

Blocks are compressed independently in a JPEG image\footnote{Only exception to that is DC values are predicted from the previous block.}
However, In H.264, a macroblock can be predicted from adjacent macroblocks of the same frame which makes the block an intra-coded block, or, it can be predicted from macroblocks of previous and/or future frames, which makes it an inter-coded block.
The difference between the original block and its prediction is called residual.
It is the residual that is being transformed, quantized, entropy coded and transferred alongside the prediction parameters.
In this way, the receiver can construct the estimate of the original block using the reference block, residual and prediction parameters.

A frame in a H.264 or H.265 video can have three types: I, B and P. An I frame is self contained and temporally independent.
It can be likened to a JPEG image in this sense.
However, they are quite different in many aspects including variable block sizes, use of prediction, integer transformation based on DCT, variable quantization, and deployment of a loop filter.
The macroblocks in P frame can use previous I or P frame macroblocks to find the best prediction. 
Encoding of a B frame provides the same flexibility. 
Moreover, future I and P frames can be utilized when predicting B frame macroblocks.

In a similar manner, macroblocks can be categorized into three types, namely, I, P and B.
An I microblock is intra-coded with no dependence on future or previous frames, and I frames only contain this type of blocks. 
A P macroblock can be predicted from a previous P or I frames, and P frames can contain both types of blocks.
Finally, a B macroblock can be predicted using one or two frames of I or P types.
A B frame can contain all three types of blocks.
A type sequence of frames, e.g., IBBPBBPBB, is called group of pictures (GOP) and this pattern repeats over the whole video.
The GOP must begin with an I frame.

\subsection{Quantization}			
					
In H.264, the transform and quantization steps are designed to minimize computational complexity so that it can be deployed by devices using limited-precision integer arithmetic.
For this purpose, instead of using discrete cosine transform (DCT) as used in JPEG, H.264 uses an integer transform.
Moreover, some parts of the integer transform is combined with quantization into a single step as follows.
Integer transform uses a scaled integer approximation of a DCT transform matrix.
However, unlike a DCT matrix, an integer transformation matrix is not orthogonal.
For orthogonalization, it is multiplied by a scaling matrix.
Later, the transformed matrix is divided by a quantization step.
Instead of having a multiplication for scaling and then a division for quantization, H.264 standard offers merging these two operations into one multiplication step which can be realized in hardware very fast.
The H.265 standard also uses the same quantization approach.
					
Because of this in H.264 standard the actual quantization step size cannot be selected by the user but rather controlled indirectly as a function of a quantization parameter.
In practice, user decides on the bitrate of coded video and the quantization parameters is varied accordingly so that the intended rate can be achieved.
Therefore, unlike a single quantization table as deployed in JPEG, the quantization parameters might change several times when coding a video.
Also, in contrast to JPEG, a uniform quantizer step is applied to every coefficient in a $ 4 \times 4 $ or $ 8 \times 8 $ block by default.
However, frequency dependent quantization can also be performed for different compression profiles.

\subsection{Loop Filtering}
Block-wise quantization performed during encoding yields a blockiness effect on the output images, H.264 decoder uses loop filters to compensate this effect it by applying a spatially variant low pass filter to smoothen the block boundaries.
The filter is applied up to 3 pixels from the boundary of $ 4 \times 4 $ blocks.
The strength of the filter and the number pixels affected from filtering depends on several constraints such as being at the boundary of a macroblock, current QP, and the gradient of image samples across the boundary.

Similar to H.264, an in-loop filtering process is applied in H.265.
However, loop filtering in H.265 has two steps.
In the first step, a deblocking filter (DBF), which is similar to the loop filter of H.264 format, is applied.
The DBF is simplified in H.265 such as there are three Boundary Strength (BS) values instead of five. 
Also, it is only applied to the edges that are aligned on an $ 8 \times 8 $ sample grid rather than $ 4 \times 4 $ sample grid. 
In the second step, DBF is followed by a Sample Adaptive Offset (SAO) filtering which does not exist in H.264 standard. The purpose of SAO is reducing the ringing artefacts and applying an additional refinement to the reconstructed image.
Applying SAO after DBF improves both the objective and subjective quality around edges and smooth areas even further \cite{tan2012objective}.




It must be noted that the loop filter is also deployed during encoding to ensure synchronization between encoder and decoder.
In essence, the decoder reconstructs each macroblock by adding a residual signal to a reference block identified by the encoder during prediction.
At the decoder, however, all the previously reconstructed blocks would have been filtered.
To not create such a discrepancy, the encoder also needs to perform prediction assuming filtered block data rather than using original macroblocks.
The only exception to this is the intra-frame prediction where extrapolated neighboring pixels are taken from an unfiltered block.

\section{Mitigation of Video Coding artefacts}

The ability to effectively cope with video coding related information loss essentially requires an intervention on the decoding stage 
during which a compressed video file is turned into a sequence of frames. 
To realize this we take a two-pronged approach.
First, we eliminate the loop filtering step at the decoder while ensuring the decoder stays in lockstep with the encoder.
Second, we retain decoding information at the macroblock level, including macroblock sizes, locations, and quantization parameters, 
for each video frame and utilize it to perform more reliable estimation. 
Details of our approach are elaborated in the following subsections. 

\subsection{Compensation of Loop Filtering}

The strength of blocking effect at block boundaries of a video frame is crucially determined by the level of compression.
In high quality videos, such an artefact will be less apparent. 
Therefore, from a PRNU estimation point of view, presence of loop filtering will be of lesser concern. 
However for low quality videos, application of loop filtering will result with removal of significant portion of PRNU pattern, thereby making source attribution less reliable \cite{tan2016video}. 

\begin{figure}[ht!]
\centering
\includegraphics[width=1\columnwidth]{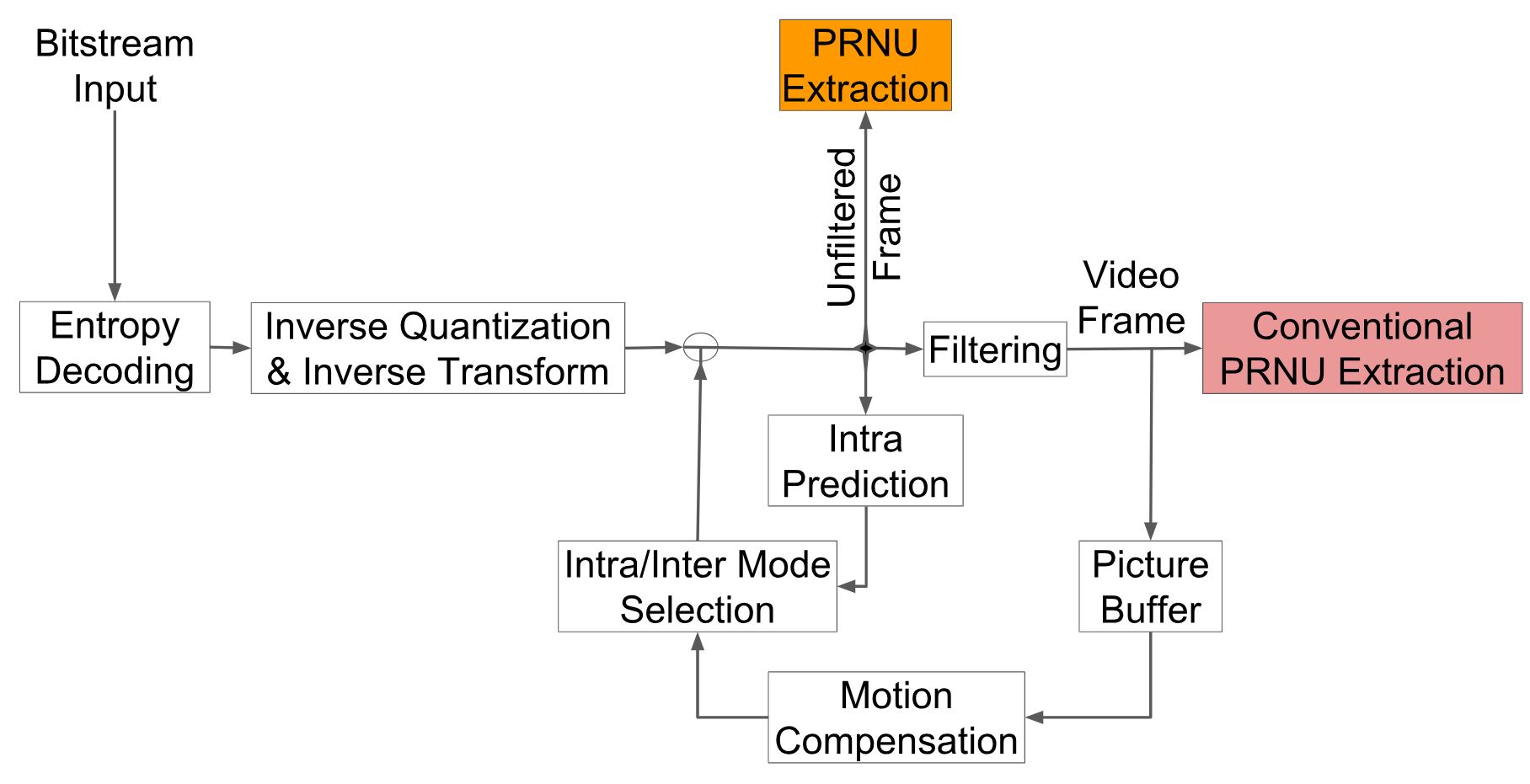}
\caption{Simplified schema of decoding for H.264 and H.265 codecs. In the block diagram, for H.264, the filtering block represents the deblocking loop filter and  for H.265, it involves a similar deblocking filter cascaded with the sample adaptive offset filter.}
\label{fig:loop}
\end{figure}

A simplified decoding schema for H.264 and H.265 codecs is given in Fig. \ref{fig:loop}.
In this model, the coded bit sequence is run through entropy decoding, dequantization, and inverse transformation steps before the residual block and motion vectors are obtained.
Finally, video frames are reconstructed through the recursive motion compensation loop. 
The loop filtering is the last processing step before visual presentation of a frame. It also weakens the PRNU pattern. Therefore, avoiding loop filtering will yield more efficient PRNU extraction.
Despite the fact that loop filter is primarily related to decoder side, it is also deployed in encoder side to be coherent with the decoder. In the encoder side, decoded picture buffer is used for storing reference frames and those pictures are all loop filtered. Moreover, when a macroblock is predicted from neighbouring ones, those referenced macroblocks are also loop filtered before the prediction. 

It can be seen in Fig. \ref{fig:loop} that for intra-prediction unfiltered blocks are used. 
Since I type macroblocks that appear in all frames are coded using intra-frame prediction; bypassing the loop filter during decoding will not introduce any complications. 
However for P and B type macroblocks since encoder performs prediction assuming filtered blocks, simply removing the loop filter at the decoder will not yield a correct reconstruction.
To compensate for this behavior that affects P and B frames, decoding process must be modified to reconstruct both a filtered and non-filtered versions of each macroblock. 
Filtered macroblocks must be used for reconstructing future macroblocks, and non-filtered ones need to be used for PRNU estimation. 
To realize this, we modified the operation of H.264 and H.265 decoding modules of open source libraries developed under the FFMPEG project with required additional steps.

As a result of this modification, video frames will exhibit blocking artefacts resembling those introduced by JPEG coding with two significant differences: the size of blocks and locations of artefacts. In JPEG compressed images, blockiness artefact positions are well known because block sizes are constant over the image. In videos, by contrast, (sub-)macroblock sizes are not deterministic. Video coding standards leave encoding preferences including block sizes to the developers \cite{sze2014high}. Block size is a trade-off between picture quality and bitrate. Smaller block sizes produce finer pictures but cost more in bitrate. This decision is typically optimized by using rate-distortion optimization \cite{sullivan2012overview}.
As a result, PRNU estimates containing blocking artefacts are very unlikely to exhibit a structure that is persistent across many consecutive frames. Therefore, we can rely on the estimation process in Eq.\ref{eq2} itself to suppress blocking artefacts through averaging.

\subsection{Coping With Quantization Related Information Loss}
For both photos and videos, the most important factor that determines the reliability of the estimated PRNU factor is the loss of information caused by quantization.
Measurements performed on diverse set of photos show that almost all camera models save their output as high quality JPEG coded images with a default quality factor in the range 85-95 \cite{uzun2015carving}; hence, compression does not pose a significant obstacle to PRNU estimation in most cases.
In contrast to photos, file size of a video is a major concern as it significantly increases the demand for storage and transmission resources. 
Therefore, cameras typically downsize high-resolution sensor output, further degrading the PRNU pattern \cite{ei2019}, and perform a heavier compression. 

To provide a relative comparison between effects of video and still image compression on PRNU estimation, we performed a test.
With JPEG coding, the strength of compression is determined by the quality factor ($QF$), a number that varies between 1 and 100 such that at a quality of 100 no quantization (other than integer rounding) is performed and higher quantization levels are associated with lower $QF$ values.
Similarly, in video coding, the extent of compression is determined in terms of the quantization step size.
In both H.264 and H.265 codecs, the value of quantization step size, $Qstep$, is indexed through a quantization parameter $QP$.
The parameter $QP$ takes values between 1 and 51 with lower values indicating that lesser amount of quantization noise is introduced to each transform coefficient. 
(The relation between the two parameters is such that $Qstep$ is 1 when $QP$ is set to 4, and for each increment of 6 in $QP$, $Qstep$ value doubles.)

Tests are performed on a set of high resolution photos ($4160 \times 3120$) acquired at a JPEG quality of 100 while slightly moving the camera in one direction. 
A reference PRNU pattern is estimated from 150 images and 120 images are set aside for comparison.  
This subset of images are first compressed at all quality values, yielding a total of $99\times 120$ images. 
The same images are then H.264 encoded (without turning off the loop filter) at fixed QP values which resulted with 51 compressed videos. 
The mean squared error (MSE) between the original and compressed images are computed and averaged over all $QF$ values. 
The same computation is performed for frames extracted from the videos to determine the average MSE corresponding to all $QP$ values. 
Then, $QP$ and $QF$ values that yield similar MSE values are identified. 
The same process is repeated by first estimating the PRNU patterns from all resulting images and frames and then
evaluating their match with the reference PRNU pattern in terms of the PCE metric.
Average PCE values at all $QP$ and $QF$ values are subsequently computed and compared.

Table \ref{jpeg2QP} identifies $QP$ and $QF$ pairs that yield similar MSE and PCE values.
It can be seen that H.264 coding, in comparison to JPEG coding, maintains a relatively high image quality (measured in the MSE sense), despite rapidly weakening the inherent PRNU pattern.
For example, compression with $QP=21$ introduces an MSE distortion that is comparable to JPEG compression at $QF=89$; however, PCE values match those of a higher compression, $QF=66$.
It is observed from these results that what would be considered medium level in video compression ($QP=20$ to $QP=30$) is equivalent to heavy JPEG compression ($QF=66$ to $QF=32$).
This increasing gap essentially indicates why conventional method of PRNU estimation is not very effective on videos.

\begin{table}[!ht]
	\centering
	\caption{A Comparative Evaluaiton of JPEG and H.264 Encoding in terms of MSE and PCE}
	\label{jpeg2QP}
	\begin{tabular}{c|c||c|c}
		\toprule
		\multicolumn{2}{c||}{Equalized MSE} & \multicolumn{2}{c}{Equalzied PCE}\\
		\hline
		\hline
		$ QF_{JPEG} $ & $ QP_{H.264} $   &  $ QF_{JPEG} $ & $ QP_{H.264} $\\
		\hline
		92 & 5  & 92 & 5 \\   \hline
	92 & 10 &   90 & 10\\  \hline
	91 & 15 &   84 & 15\\  \hline
	90 & 18 &   77 & 18\\  \hline
	89 & 21 &   66 & 21\\  \hline
	87 & 24 &   50 & 24\\  \hline
	82 & 27 &   33 & 27\\  \hline
	77 & 30 &   22 & 30\\  \hline
	67 & 35 &   10 & 35\\  \hline
	61 & 40 &   3 & 40\\  \hline
	57 & 45 &   1 & 45\\  \hline
	57 & 50 &   1 & 50\\  \hline

        \bottomrule
	\end{tabular}
\end{table}



In many video coding scenarios, however, user decides on the bitrate of coded video and the encoder changes the value of $QP$ adaptively to attain the target bitrate. 
Further, unlike JPEG compression where a single quantization table is deployed for compression of all $8\times8$ image blocks, $QP$ parameter might change for each block when coding a frame.
As a result, the effect of compression on the reliability of the estimated PRNU pattern depends on the video content itself and the designated level of compression, and devising a parametric relation between the two is a challenging task. 


The effect of compression on PRNU estimation can be empirically determined in terms of the quantization parameter $QP$ and the PCE metric.
Due to the exponential relation between $QP$ and $Qstep$, a linear increase in $QP$ will cause a
similar degradation in quality measured in terms peak signal-to-noise ratio. 
Hence, it can be expected that the strength of estimated PRNU will also decrease exponentially with increasing $QP$.
To test the validity of this assumption, we performed tests on a set of videos captured by 25 different smartphone cameras.
These videos were captured under controlled settings by turning off stabilization and electronic zoom and at the highest  frame resolution supported by the camera with the best video bitrate possible (corresponding to a $QP$ value of 1) using the camera app developed in \cite{ei2019}. 

\begin{figure}[ht!]
\centering
\includegraphics[width=0.8\columnwidth]{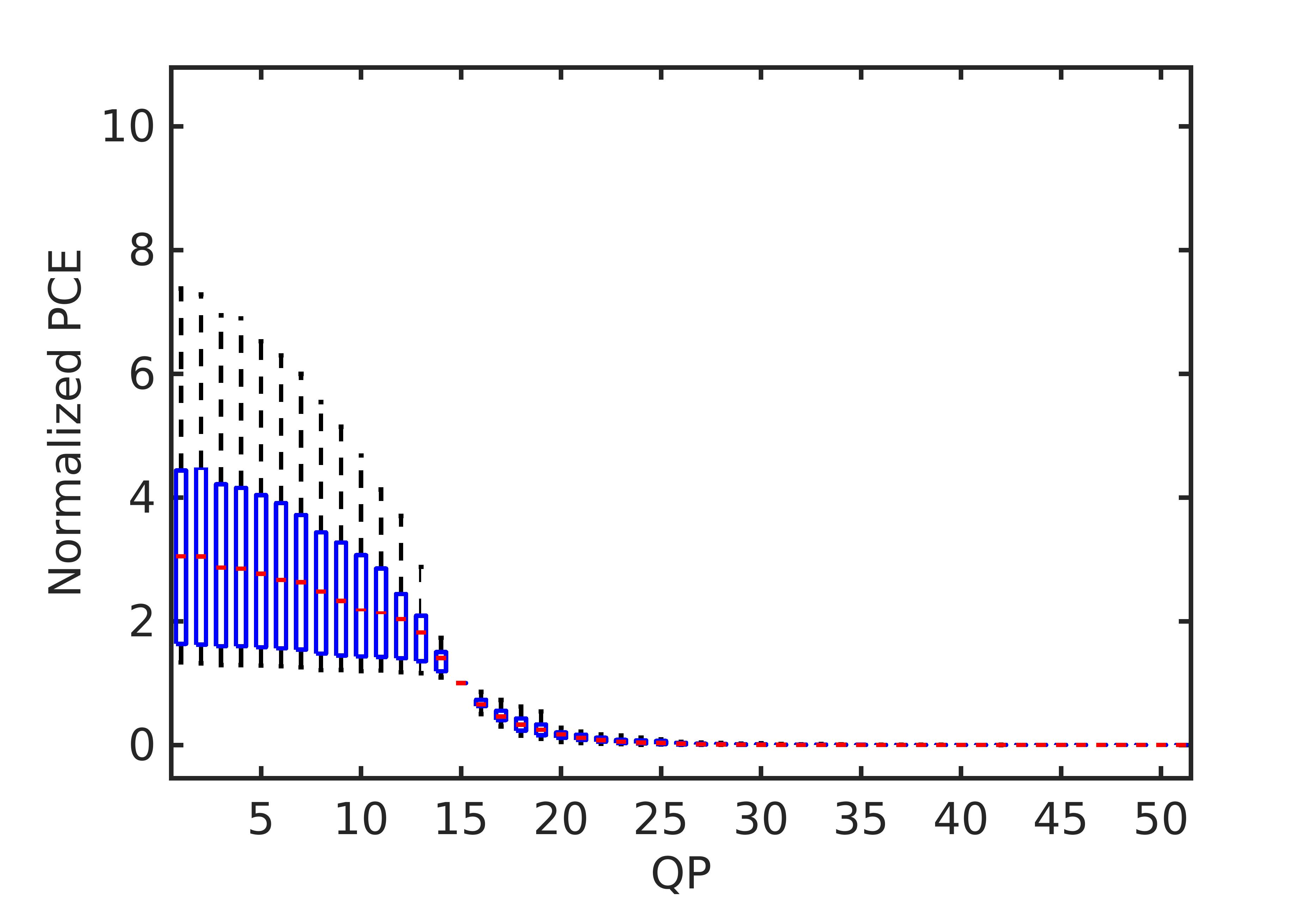}
\caption{Change in PCE values with respect to $QP$ computed using 25 videos captured by different cameras. 
PCE values are normalized with respect to the PCE value at $QP=15$ for more compact representation of within-camera variation.}
\label{fig:QP}
\end{figure}

Using two videos from each camera, we performed the following tasks:
{\em (i)} camera sensor's reference PRNU pattern is obtained using one of the videos; 
{\em (ii)} the other video is compressed at all $QP$ values, yielding a total of 51 compressed versions of the same video;  
{\em (iii)} resulting videos are decoded and individual PRNU patterns are estimated from all frames;
{\em (iv)} extracted PRNU patterns are matched with the reference pattern and an average PCE value is computed for all $QP$ values; and 
{\em (v)}  average PCE values are normalized with respect to the average PCE value at $QP=15$ to further reduce the variation among cameras.
Figure \ref{fig:QP} displays the change in PCE with respect to $QP$ by combining results from 25 cameras in the box-plot format. 
It can be seen that PCE values show a slow, almost linear, initial decline until $QP$ is around 10 followed by an exponential drop for increasing values
with PRNU pattern getting completely removed when $QP$ goes beyond 28.
We consider two strategies to cope with this behavior.


\subsubsection{PRNU masking}

Parts of PRNU pattern estimated from macroblocks that are compressed at higher $QP$ values will only yield noise. 
Therefore, those macroblocks should be excluded from the estimation process.  
To incorporate this into the estimation process we create a pixel-wise mask, $M$, for each frame as it is being decoded. 
In the mask $M$, all pixel locations corresponding to macroblocks that underwent compression with $QP$ higher than 28 are set to 0, and the rest of the pixel values are set to 1.
The PRNU factor $K$ is estimated using Eq. \ref{eq2} by multiplying each contributing frame by this mask as
\begin{equation}
K=\dfrac{\sum_{i=1}^{N} I_{i} \times W_{i} \times M_{i} }{\sum_{i=1}^{N} (I_{i})^2 \times  M_{i}}
\label{masking}
\end{equation}

\begin{figure}[ht!]
\centering
\includegraphics[width=1\columnwidth]{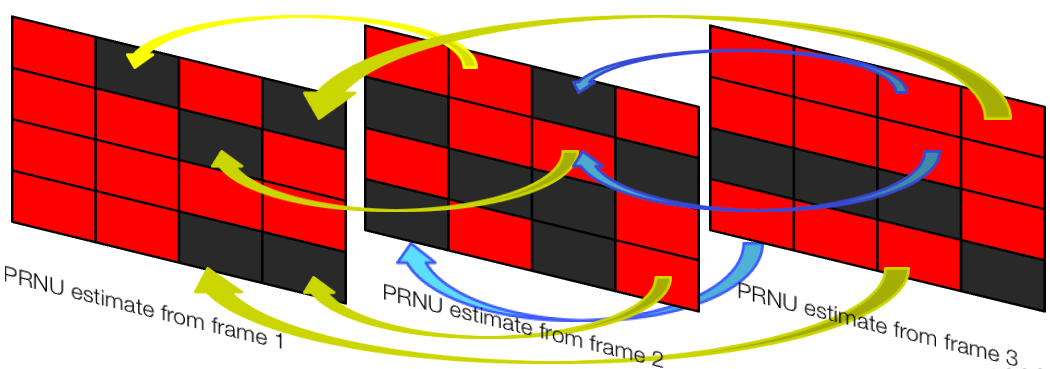}
\caption{Rearrangement of remaining macroblocks after binary masking of each frame.  
Black colored blocks represent the locations of masked out macroblocks and arrows show the order of splicing depending on factors such as $QF$, intensity, texture, etc.} 
\label{fig:splicing}
\end{figure}

The use of binary mask to eliminate highly compressed blocks can be further enhanced by sorting macroblocks in order of decreasing reliability \cite{eusipco18}.
In other words, to maximize PCE, we can fill up new frames by select PRNU noise blocks from other frames as shown in Fig. \ref{fig:splicing}. 
When splicing together new frames of PRNU patterns, macroblocks can be rearranged with respect to their $QP$ values
as well as content characteristics, such as intensity levels and texture, that are known to affect PRNU estimation. 
It must be noted that in the newly generated frames, it is necessary for macroblocks to preserve their position in the original frames to prevent any geometric distortion.
Resulting spliced-up frames can then either be used to match with the reference PRNU pattern or to create of more reliable PRNU estimates.

\subsubsection{PRNU Weighting}

\begin{figure}[ht!]
\centering
\includegraphics[width=0.65\columnwidth]{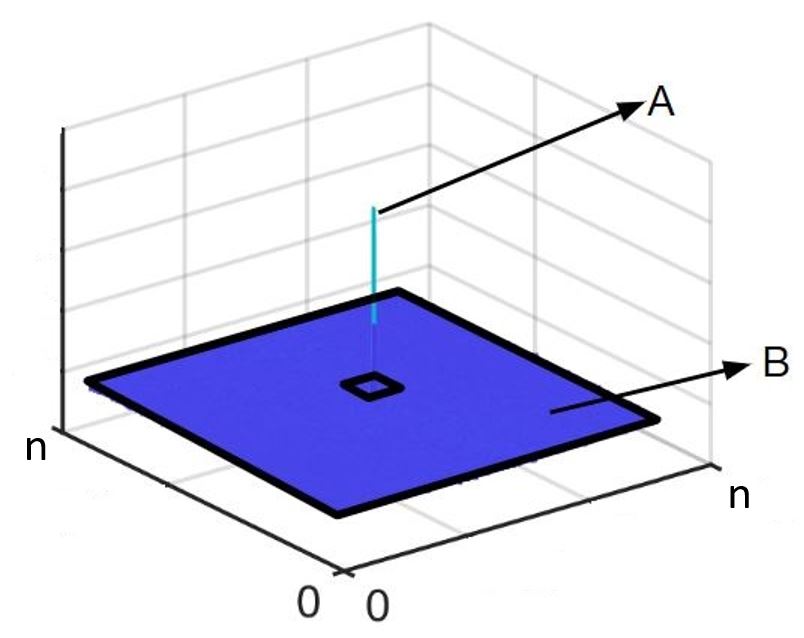}
\caption{PCE is calculated as the square of the ratio of cross-correlation term at A to the average of square terms in B {\em i.e.}, $PCE=\frac{A^2}{(mean(B^2))}$}
\label{fig:crosscorr}
\end{figure}

An alternative approach to binary masking is to weight each macroblock depending on the level of compression
so that blocks that underwent lighter compression contributes more strongly to estimating the PRNU pattern. 
Having empirically determined the relation between $QP$ and PCE as shown in Fig. \ref{fig:QP}, we can utilize it to determine appropriate weighting factors. 
For the general case, given two PRNU factor estimates $K$ and $K^{*}$, the PCE is defined as 
the ratio between the square of correlation between $K$ and $K^*$ and the total energy in the cross-correlation plane, which can be calculated as:
\begin{equation}
\label{NCCC}
\begin{aligned}
c(k,l)=\frac{1}{n^2} \sum_{i=1,j=1}^{n,n} K_{i,j} \cdot K_{i \oplus k,j \oplus l }^{*},\,\,   k,j= 0,1,\ldots n-1 
\end{aligned}
\end{equation}
\begin{equation}
\label{pce}
\begin{aligned}
PCE(K,K^{*})=\frac{c^2(0,0)}{\frac{1}{n^2-\|\delta\|} \sum_{i,j,(i,j)\notin\delta}c^2(k,l)}
\end{aligned}
\end{equation}
where $\oplus$ denotes a modulo n addition to compute cyclically shifted cross-correlation 
within a window of size $n\times n$ except for a $\delta$ region around the correlation peak.
The terms involved in computation of PCE is depicted pictorially in Fig. \ref{fig:crosscorr}.

\begin{figure}[ht!]
\centering
\includegraphics[width=0.7\columnwidth]{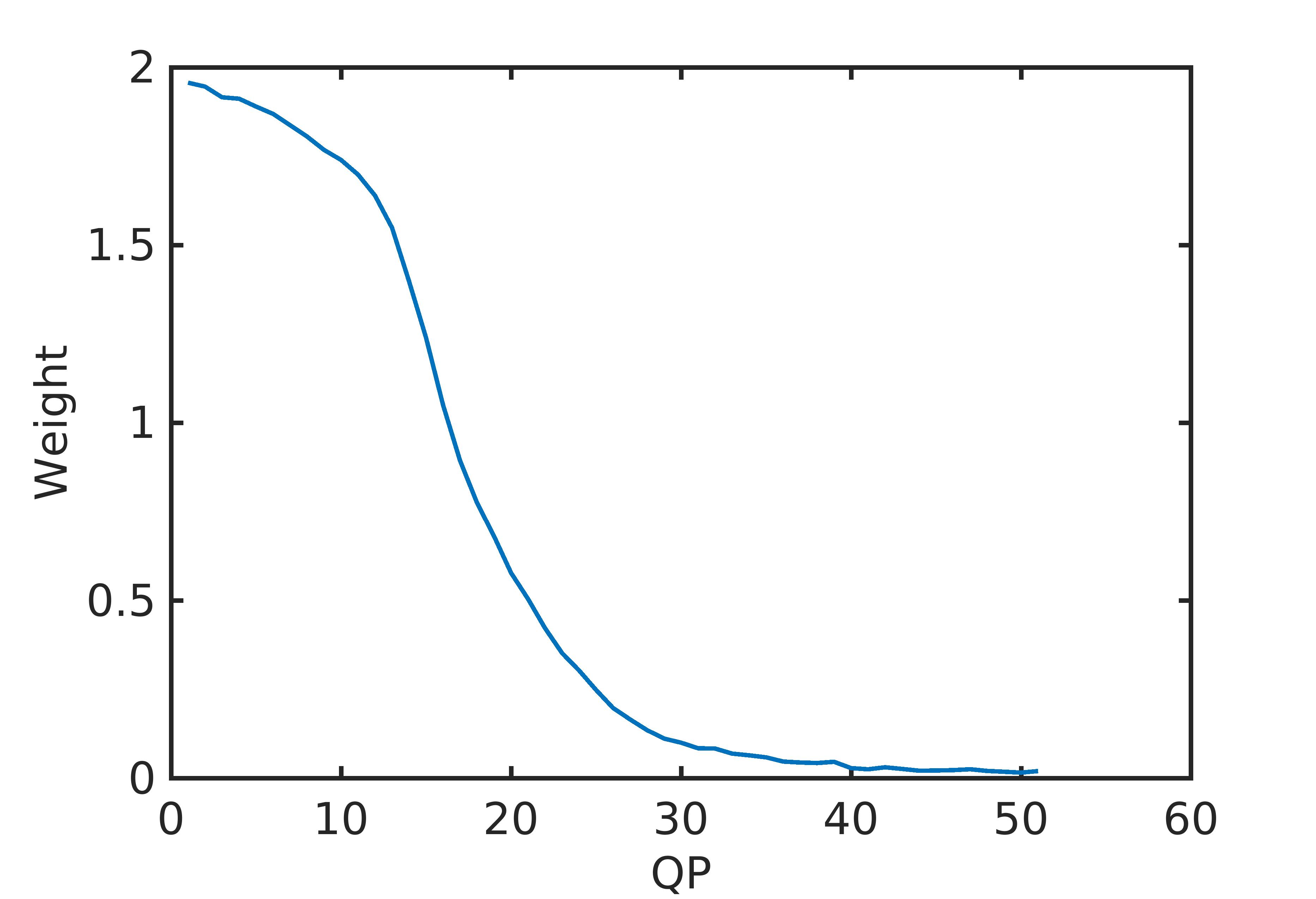}
\caption{The weight curve as a function of QP obtained through averaging PCE-PQ curves of 25 cameras given in Fig. \ref{fig:QP}.}
\label{fig:weight}
\end{figure}

When the PRNU estimates $K$ and $K^{*}$ correspond to the same sensor, the numerator in Eq. (\ref{pce}) can be interpreted as the signal, and the denominator as the noise part since PRNU pattern is known to be pixel-wise independent.
Hence, given two PRNU instances $K^x$ and $K^y$ estimated from different media captured by a camera with reference pattern $K^{*}$, the ratio $r$ of two PCE values, {\em i.e.}, $r=\frac{PCE(K^x,K^{*})} {PCE(K^y,K^{*})}$, will be determined by the numerator of Eq. \ref{pce} as the denominator term depends on average correlation of two uncorrelated signals, which will be the same for both $K^1$ and $K^2$. 
Correspondingly, 
\begin{equation}
\sqrt{r}=\frac{c_x(0,0)}{c_y(0,0)}    
\end{equation}
where $c_i(0,0)$ corresponds to correlation between $K^i$ and $K^{*}$ as defined in Eq. \ref{NCCC}.
Assuming a linear relation between $K^x$ and $K^y$, this yields $K^x\approx \sqrt{r}K^y$.
Although PRNU patterns estimated from media at multiple compression levels will not be linearly related, this model can nevertheless be used to model the change in PCE with respect to the compression level.
This also means that given the PCE-$QP$ relation, the relative weighting factors for each $QP$ level can be determined  through the square-root of the underlying PCE-$QP$ curve normalized with respect to a base PCE value. 
Hence, the estimation process can be changed by modifying the binary values in mask $M$ introduced in Eq. \ref{masking} by the weighting factors corresponding to QP of each macroblock.
When a such a curve is not available for a camera, the general characteristic given in Fig. \ref{fig:QP} can be utilized for this purpose. 
Figure \ref{fig:weight} displays the weight function corresponding to average PCE-$QP$ curves of 25 cameras. 
Accordingly, macroblocks that underwent compression at $QP=10$ and $QP=25$ will be weighted by factors of $1.74$ and $0,25$, respectively, indicating that the former macroblocks will contribute 7 times more strongly to estimation process than the latter ones. 


\subsection{Choice of Frame Type}
The type of encoded picture (I, P, or B) is another attribute that needs to be considered when performing estimation.
Earlier work observed that use of I frames alone for PRNU estimation yields more reliable results \cite{IPBdiff}.
This was a revelation that I frames are compressed at lower compression levels than P and B frames. 
In fact, I frames are encoded by intra-frame prediction where blocks are only predicted from neighboring 
blocks within the same frame. 
On the contrary, P and B frames are coded with inter-prediction which leverages a set of reference frames to perform block predictions.
As a result of the extended search space, P and B frame block predictions result with better matching blocks.  
In other words, the prediction error that will be subject to quantization will be of lesser strength for P and B frames.
Therefore, at the same compression level with I frames, P and B frames should be expected to produce more reliable PRNU patterns. 

\begin{figure}[ht!]
\centering
\includegraphics[width=0.7\columnwidth]{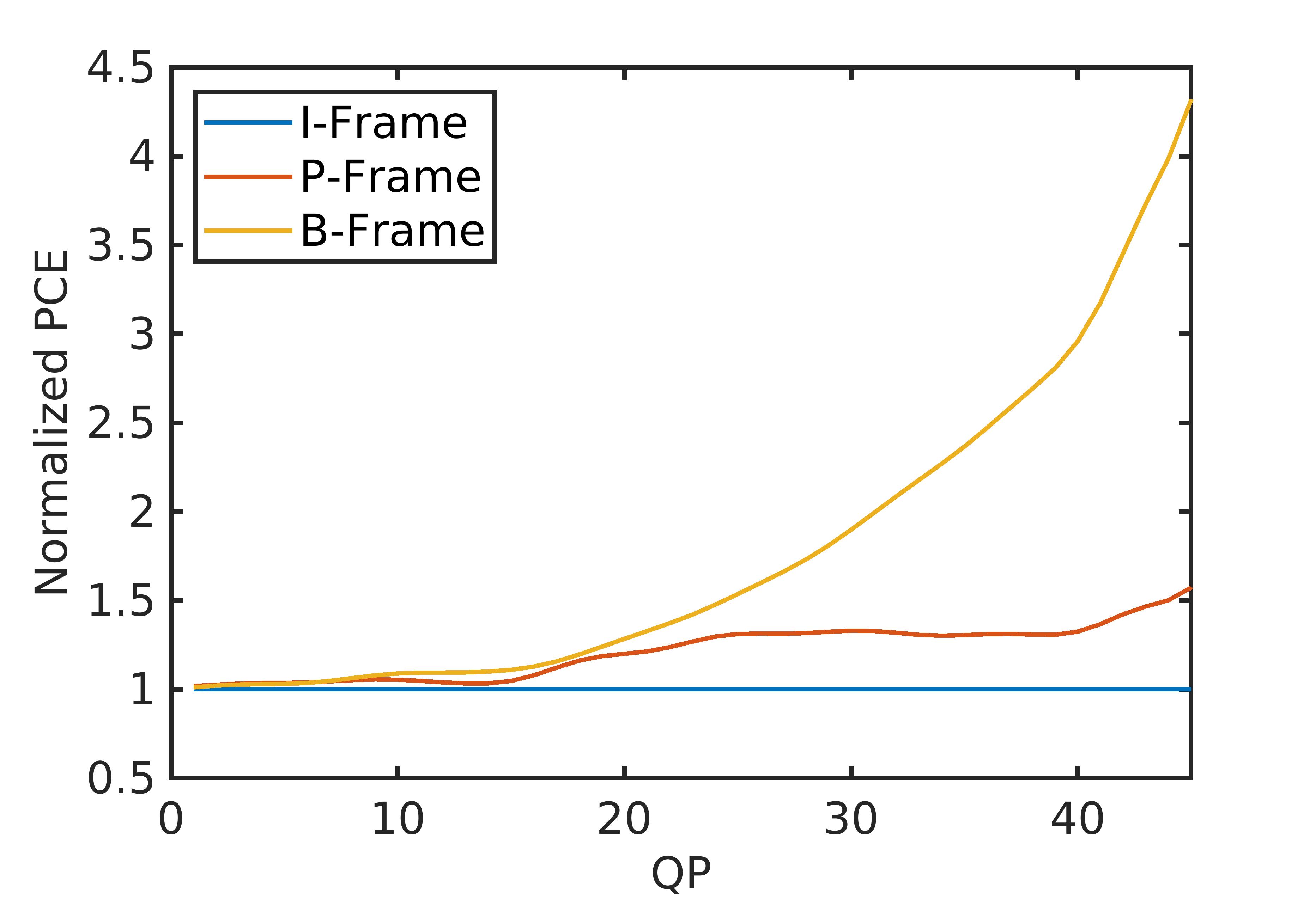}
\caption{PCE-$QP$ curves for three picture types. PCE values are normalized with respect to mean PCE values of I frames at each compression level.}
\label{fig:IPBcmp}
\end{figure}

To test the dependency of PRNU estimation on the choice of frame type, we used two raw (almost uncompressed) videos captured by two cameras. 
Videos are compressed at all $QP$ levels while setting the GOP size to 3 so that videos comprise sequences of I, P, and B frames. 
Following compression, frames are extracted and grouped based on the picture type. 
For all $QP$ levels and picture types, mean PCE was calculated and results were normalized with respect to mean PCE of I frames. 
Figure \ref{fig:IPBcmp} shows the resulting PCE-$QP$ curves for the three frame types.
Accordingly, B frames yield the best estimates with PCE values upto 4,5 times higher than those of I frames, and
P frames are slightly better than I frames at all compression levels.  
However, the difference becomes more apparent only at high compression levels where PRNU estimates are in general not very reliable as found in Fig. \ref{fig:QP}. 
Overall, these findings verify the intuition that at the same compression level B and P frames yield slightly better estimations than I frames. 
But the impact of picture type on PRNU estimation is only marginal at lower $QP$ levels; therefore, we didn't incorporate picture type as another weighting factor when generating a mask for a frame. 

\section{Experimental Results}

To determine the effectiveness of introduced approaches in countering video compression artefacts, we performed variety of tests using different datasets.
First, we compare the performance of basic PRNU estimation method with its improved versions by successively incorporating loop filter compensation and PRNU weighting methods into it. 
The dataset used for this test included videos captured by 28 Android phone cameras that support H.264 compression using our custom camera application \cite{ei2019}.
For each PRNU estimation method, cameras' reference PRNU patterns were obtained from 4 seconds long, high-quality videos. 
Another set of videos compressed at a relatively low bitrate of 2.2 Mbps were used for testing.
The test videos were also limited to 4 seconds in duration, recorded at a frame rate of 25 fps with resolutions of $2160\times2160$, $1920\times1080 $ or $1440\times1080$. 
Stabilization and electronic zoom were turned off in all cases. 
Table \ref{coded} provides average PCE values computed between the reference pattern and estimated PRNU patterns from all video frames. 
It can be seen from these results that turning off loop filter improves measured PCE values on average $2.6$ times over the basic PRNU estimation method. 
For videos that yield low PCE values with the basic method, the gain is much more significant with a $3.58$ times average improvement in PCE values.
Incorporation of masking further improves results in all videos (except for video \#4) with the PRNU weighting yielding the best results as expected.
The average improvement due to the PRNU weighting over loop filter compensation is found to be $37$\% and it is $4.44$ times over the basic estimation method.  
Similarly, for videos with low PCE values the improvement is more notable with a $5.76$ times average incease.

\begin{table}[!ht]
	\centering
	\caption{Average PCE Values for H.264 Coded Videos at 2.2 Mbps}
	\label{coded}
	\begin{tabular}{l|c|c|c|c}
		\toprule
		Video  &Basic&Loop filter &Binary&PRNU\\
		Number & method & compensation &  masking & weighting\\
		\hline
1&0.29&7.45&13.49&13.49\\   \hline
2&3.57&19.92&20.32&21.82\\   \hline
3&4.82&21.96&20.54&23.02\\   \hline
4&7.89&30.06&21.90&23.12\\   \hline
5&11.26&36.29&52.60&52.87\\   \hline
6&12.71&37.37&58.57&67.51\\   \hline
7&15.94&50.47&71.78&78.64\\   \hline
8&16.47&61.92&89.08&94.35\\   \hline
9&21.65&73.68&95.54&95.54\\   \hline
10&25.80&76.03&109.00&109.97\\   \hline
11&30.19&77.37&123.24&120.28\\   \hline
12&37.70&94.73&128.02&127.14\\   \hline
13&39.02&96.72&172.93&169.61\\   \hline
14&43.14&99.72&185.90&182.94\\   \hline
15&43.86&147.85&193.17&215.21\\   \hline
16&45.30&149.21&229.53&290.86\\   \hline
17&59.32&162.20&275.75&294.72\\   \hline
18&68.70&280.65&280.87&316.11\\   \hline
19&117.00&407.69&497.06&497.06\\   \hline
20&137.08&526.05&653.29&627.28\\   \hline
21&174.70&531.41&862.23&902.40\\   \hline
22&226.10&598.79&1048.75&1095.86\\   \hline
23&887.89&1694.85&1596.02&1673.50\\   \hline
24&5264.28&5810.72&5708.73&5569.37\\   \hline
25&5969.35&6085.48&5829.71&5807.59\\   \hline
26&7993.16&8806.05&9498.83&8792.59\\   \hline
27&30930.74&34308.76&33850.71&34271.31\\   \hline
28&97827.36&99403.07&98487.64&99108.96\\   \hline

        \bottomrule
	\end{tabular}
\end{table}

The second test involved 36 videos captured at 480p and 720p resolutions by the same 28 cameras.
(If a camera did not support 720p resolution, videos were captured only at 480p resolution.)
The videos were captured under similar controlled settings with the lightest in-camera compression ({\em i.e.}, $QP$=1).
These videos were considered as raw and were re-encoded at 12 different bitrates using both H.264 and H.265 codecs.
PRNU estimation is performed similar to above using the four methods.
Obtained average PCE values for the two codecs are given in Tables \ref{reCodedH264} and \ref{reCodedH.265}.
It can be seen that loop filter compensation and PRNU weighting improves measured results by 53\% for H.264 coding
and by 68\% for H.265 coding with more visible improvements at lower bitrates.
Tables \ref{reCodedH264per} and \ref{reCodedH.265pixel} provide similar results normalized with respect to frame resolution so that compression results can be evaluated in bits per pixel. 
It must be noted that, with binary masking, at low bitrates many macroblocks undergo heavy compression and get masked out; therefore, 
not enough macroblocks are left to perform reliable estimation.


\begin{table}[!ht]
	\centering
	\caption{Average PCE values for H.264 Encoded Videos at Different Bitrates} 

	\label{reCodedH264}
	\begin{tabular}{l|c|c|c|c}
		\toprule
		Bitrate  &Basic&Loop filter &Binary&PRNU\\
	     (kbps) & method & compensation & masking & weighting\\
		\hline
500&9.13&14.80&9.32&16.04\\   \hline
600&21.35&29.18&28.83&38.79\\   \hline
700&44.59&55.38&54.19&70.85\\   \hline
800&81.53&91.37&92.10&125.19\\   \hline
900&196.68&218.50&222.69&317.04\\   \hline
1200&371.62&414.58&424.03&615.46\\   \hline
1500&615.72&685.40&697.63&1000.42\\   \hline
2000&1334.34&1449.22&1475.15&2034.49\\   \hline
2500&2303.20&2454.78&2491.18&3286.78\\   \hline
3000&3344.96&3525.79&3580.69&4553.33\\   \hline
3500&4422.24&4580.90&4648.92&5736.10\\   \hline
4000&5492.46&5606.38&5690.03&6806.57\\   \hline

        \bottomrule
	\end{tabular}
\end{table}

\begin{table}[!ht]
	\centering
	\caption{Average PCE values for H.265 Encoded Videos at Different Bitrates}
	\label{reCodedH.265}
	\begin{tabular}{l|c|c|c|c}
		\toprule
		Bitrate  &Basic&Loop filter &Binary&PRNU\\
	     (kbps) & method & compensation & masking & weighting\\
		\hline

500&13.00&16.00&19.88&27.17\\   \hline
600&31.07&40.00&46.55&64.63\\   \hline
700&51.19&65.00&72.37&101.12\\   \hline
800&92.13&110.00&117.22&169.65\\   \hline
900&207.19&265.00&275.72&404.59\\   \hline
1200&390.53&474.00&492.69&692.14\\   \hline
1500&644.38&774.00&802.67&1101.97\\   \hline
2000&1269.84&1458.00&1489.93&1967.94\\   \hline
2500&1957.51&2178.00&2219.38&2811.35\\   \hline
3000&2771.96&3001.00&3043.66&3702.90\\   \hline
3500&3512.06&3737.00&3787.30&4461.95\\   \hline
4000&4281.75&4460.00&4515.75&5193.61\\   \hline

        \bottomrule
	\end{tabular}
\end{table}

\begin{table}[!ht]
	\centering
	\caption{Average PCE values for H.264 Encoded Videos at Different Compression Ratios in Bits/Pixels}
	\label{reCodedH264per}
	\begin{tabular}{l|c|c|c|c}
		\toprule
		Bits per &Basic&Loop filter&Binary&PRNU\\
		pixels & method & compensation & masking & weighting\\
		\hline
0.78&38.68&49.52&56.67&55.32\\   \hline
1.22&135.58&134.09&157.79&156.69\\   \hline
1.65&165.86&170.70&219.87&229.31\\   \hline
2.00&208.25&225.98&292.23&308.37\\   \hline
2.38&166.33&225.10&279.43&309.15\\   \hline
2.80&581.71&649.89&867.65&986.65\\   \hline
3.55&1418.41&1554.66&1977.09&2229.94\\   \hline
4.47&1822.74&1985.70&2470.40&2742.85\\   \hline
5.52&3804.83&4014.48&4856.10&5308.20\\   \hline
7.27&1967.53&2076.29&2501.48&2728.98\\   \hline
9.57&3834.22&3961.46&4573.40&4853.24\\   \hline
12.41&4946.73&4945.57&5480.46&5726.32\\   \hline

        \bottomrule
	\end{tabular}
\end{table}

\begin{table}[!ht]
	\centering
	\caption{Average PCE values for H.265 Encoded Videos at Different Compression Rations in Bits/Pixels}
	\label{reCodedH.265pixel}
	\begin{tabular}{l|c|c|c|c}
		\toprule
		Bits per &Basic&Loop filter &Binary&PRNU\\
		pixels & method & compensation & masking & weighting\\
		\hline

0.78&51.15&64.98&85.40&80.56\\   \hline
1.22&124.65&118.12&154.18&212.66\\   \hline
1.65&179.06&141.93&182.15&311.51\\   \hline
2.00&263.21&223.72&277.21&445.41\\   \hline
2.38&235.78&205.16&250.77&397.58\\   \hline
2.80&654.66&433.11&529.70&1169.21\\   \hline
3.55&1362.87&781.27&921.71&2210.38\\   \hline
4.47&1692.64&1111.61&1286.95&2583.76\\   \hline
5.52&3246.11&2454.14&2736.32&4527.24\\   \hline
7.27&1533.15&1309.83&1513.50&2071.75\\   \hline
9.57&2974.39&2630.98&2898.97&3567.46\\   \hline
12.41&3665.65&3052.18&3280.88&4062.84\\   \hline

        \bottomrule
	\end{tabular}
\end{table}


Another important evaluation criteria is the length of the minimum duration video needed to reliably identify the source camera. 
This is a major concern in practice when large number of videos need to be processed and computational resources are limited.
To determine this, we used 21, 30-second long videos captured at 15 Mbps bitrate under controlled settings.  
As earlier, videos are re-encoded at 12 bitrates using H.264 and H.265 codecs. 
Using all the methods other than the binary masking, we estimate the PRNU pattern from all accumulated frames at the end of each second until the 
the match with the reference PRNU pattern yields a PCE value of 60, the typical threshold used for verifying the source of photos.
For each method we obtained the average duration in seconds required to achieve the target PCE threhold value.
Obtained average times are presented in Table~\ref{reCodedH264second} and Table~\ref{reCodedH.265second} for H.264 and H.265 encoding, respectively. 
It can be seen that for low bitrate videos (500-2,500 Kbps), the two methods combined reduce the required duration to 35\% and 38\% of what is needed by the basic method, respectively, for H.264 and H.265.



\begin{table}[!ht]
	\centering
	\caption{Average Duration of a H.264 Coded Video Needed for Reliable Match (seconds)} 
	\label{reCodedH264second}
	\begin{tabular}{l|c|c|c}
		\toprule
		Bits per &Basic&Loop filter&PRNU\\
		pixels & method & compensation  & weighting\\
		\hline

500&14.18&12.45&4.59\\ \hline
600&12.82&11.27&3.95\\ \hline
700&9.95&8.32&3.32\\ \hline
800&8.91&6.95&3.14\\ \hline
900&7.00&6.09&2.55\\ \hline
1200&4.14&4.23&1.91\\ \hline
1500&3.36&3.14&1.32\\ \hline
2000&2.41&2.14&1.09\\ \hline
2500&1.77&1.77&0.91\\ \hline
3000&1.59&1.55&0.95\\ \hline
3500&1.41&1.14&0.95\\ \hline
4000&1.27&1.27&0.91\\ \hline

        \bottomrule
	\end{tabular}
\end{table}


\begin{table}[!ht]
	\centering
	\caption{Average Duration of a H.265 Coded Video Needed for Reliable Match (seconds)} 
	\label{reCodedH.265second}
	\begin{tabular}{l|c|c|c}
		\toprule
		Bits per &Basic&Loop filter&PRNU\\
		pixels & method & compensation  & weighting\\
		\hline
		
500&14.77&12.00&5.59\\ \hline
600&12.64&9.68&4.36\\ \hline
700&8.73&7.18&3.36\\ \hline
800&8.00&5.68&2.82\\ \hline
900&6.05&4.41&2.09\\ \hline
1200&3.64&3.05&1.50\\ \hline
1500&2.64&2.00&1.27\\ \hline
2000&1.77&1.64&0.91\\ \hline
2500&1.64&1.64&0.91\\ \hline
3000&1.36&1.23&0.77\\ \hline
3500&1.23&1.18&0.68\\ \hline
4000&1.05&0.86&0.64\\ \hline

        \bottomrule
	\end{tabular}
\end{table}


\section{Conclusions}

Estimation of PRNU pattern of a sensor from a video involves many challenges. 
The change in the sensor’s acquisition behavior, the necessity for stabilization, and the need for very effective compression are at the core of these challenges.
Therefore, PRNU estimation and verification procedures need to be adapted to mitigate against disruptive effects of such processing steps in the camera pipeline.
In this work, we introduced methods to tackle artefacts related to widely deployed H.264 and H.265 video coding standards. 

Our analysis demonstrate that the use of the deblocking filter and the quantization related information loss impair PRNU pattern significantly at medium to high video compression levels. 
To cope with effects of these encoding steps, we took a proactive approach by modifying the decoding process and utilizing block-level encoding parameters to 
weigh the contribution of each macroblock in accordance with the level of compression it underwent.
Results show that incorporation of the introduced loop filter compensation and PRNU weigthing methods into the conventional estimation method yields on average $3-5$ times increase in measured PCE values. 

Our results further show that at the same compression level P and B type frames yield more reliable PRNU estimates than I frames due to a more accurate prediction which in turn introduces lesser quantization noise. 
Our findings also implicitly verify the superiority of H.265 encoding over H.264 in preserving image quality at lower compression levels. 
It is observed that at low to medium bitrates, H.265 coded videos yield higher PCE values than H.264 coded videos.
In contrast, at higher bitrates, this trend switches which indicates that H.264 is less intrusive to content in that compression regime.

The performance gain due to use of proposed methods can also be translated into time complexity gains in PRNU estimation as they allow for using shorter duration videos for source attribution.
This latter aspect also contributes to reduction of costs associated with downloading and storage of long videos especially when large number of videos have to be processed.
Finally, the ability to mitigate video coding artefacts marks a vital step in devising effective methods for stabilized video.

\vspace{-0.3cm}
\section{Acknowledgement}
This work is supported by the Scientific and Technological Research Council of Turkey (TUBITAK) grant 116E273.

\bibliographystyle{IEEEtran}
\bibliography{bibfile}

\end{document}